\documentclass{aa}  

\usepackage{graphicx,txfonts,lscape,natbib}
\bibpunct{(}{)}{;}{a}{}{,} 

\begin{document}
   \title{A search for lithium in metal-poor L dwarfs
}


   \author{N.\ Lodieu \inst{1,2}\thanks{Based on observations collected at the European Southern Observatory, Chile, under programmes 089.C-0883 and 091.C-0594A}\thanks{Based on observations made with the Gran Telescopio Canarias (GTC), installed in the Spanish Observatorio del Roque de los Muchachos of the Instituto de Astrof\'isica de Canarias, in the island of La Palma (programmes GTC64\_10B and GTC38\_11A)},
          A.\ J.\ Burgasser \inst{1,2,3,4},
          Y.\ Pavlenko \inst{5,6},
          R.\ Rebolo \inst{1,2,7}
          }

   \institute{Instituto de Astrof\'isica de Canarias (IAC), Calle V\'ia L\'actea s/n, E-38200 La Laguna, Tenerife, Spain. 
         \email{nlodieu,rrl@iac.es}
         \and
         Departamento de Astrof\'isica, Universidad de La Laguna (ULL), E-38205 La Laguna, Tenerife, Spain.
         \and
         Center for Astrophysics and Space Science, University of California San Diego, La Jolla, CA 92093, USA
         \and
         Visiting professor at the Instituto de Astrof\'isica de Canarias (IAC), La Laguna, Tenerife, Spain
         \and
         Main Astronomical Observatory of the National Academy of Sciences of Ukraine.
         \and
         Center for Astrophysics Research, University of Hertfordshire, College Lane, Hatfield, Hertfordshire AL10 9AB, UK
         \and
         Consejo Superior de Investigaciones Cient\'ificas, CSIC, Spain.
             }

   \date{Received \today{}; accepted \today{}}
 
  \abstract
   {}
   {The aim of the project is to search for lithium in absorption at 6707.8\AA{}
to constrain the nature and the mass of the brightest low-metallicity L-type dwarfs 
(refered to as L subdwarfs) identified in large-scale surveys.}
   {We obtained low- to intermediate-resolution (R\,$\sim$\,2500--9000) optical 
($\sim$560--770 nm) spectra of two mid-L subdwarfs, SDSS\,J125637.13$-$022452.4 
(SDSS1256; sdL3.5) and 2MASS\,J162620.14$+$392519.5 (2MASS1626; sdL4) with spectrographs 
on the European Southern Observatory Very Large Telescope and the Gran Telescopio de Canarias.
}
   {
We report the presence of a feature at the nominal position of the lithium absorption 
doublet at 6707.8\AA{} in the spectrum of SDSS1256, with an equivalent width of 
66$\pm$27\AA{} at 2.4$\sigma$, which we identify as arising from a CaH molecular
transition based on atmosphere models. We do not see any feature at the position of 
the lithium feature in the spectrum of 2MASS1626\@. The existence of overlapping molecular
absorption sets a confusion detection limit of [Li/H]\,=\,$-$3 for equivalently-typed
L subdwarfs. We provided improved radial velocity measurements of 
$-$126$\pm$10 km~s$^{-1}$ and $-$239$\pm$12 km~s$^{-1}$ for SDSS1256 and 2MASS1626,
respectively, as well as revised Galactic orbits. 
We implemented adjusting factors for the CaH molecule in combination with the
NextGen atmosphere models to fit the optical spectrum of SDSS1256 in the 6200--7300\AA{} range.
We also estimate the expected Li abundance from interstellar accretion ([Li/H]\,=\,$-$5),
place limits on circumstellar accretion (10$^{9}$ g/yr), 
and discuss the prospects of Li searches in cooler L and T subdwarfs.
}
   {}


  \authorrunning{Lodieu et al$.$}
  \titlerunning{Lithium in metal-poor L dwarfs}

   \maketitle
%

%
%
\section{Introduction}
\label{Lithium_sdL:intro}

Stars spend most of their lifetime on the main-sequence in hydrostatic
equilibrium, burning hydrogen. To the contrary, brown dwarfs never reach 
core temperatures and pressures high enough to fuse hydrogen \citep{kumar63b,hayashi63}.
Lithium (Li) is destroyed in stellar interiors via collisions with protons. 
At solar metallicities brown dwarfs with masses below 0.065 M$_{\odot}$ do 
not reach the temperatures needed for Li\,I burning \citep{magazzu91,rebolo92,basri96}. 
The precise mass below which Li\,I is preserved depends on metallicity and, 
according to models, it could be higher at lower metallicities 
\citep{burrows93,chabrier97,bildsten97,baraffe98,chabrier00a,burke04}.

Low-resolution optical spectroscopy of solar-metallicity nearby L dwarfs suggests
early-L dwarfs represent a mixture of low-mass stars and brown dwarfs whereas
most objects with spectral types later than $\sim$L5 are substellar.
From a large sample of nearby L dwarfs with optical spectra, \citet{kirkpatrick08}
showed that 10--20\% of L0--L2 dwarfs exhibit Li in absorption, the percentage 
increasing to 40--80\% for mid- to late-L dwarfs. This trend is corroborated 
by other works, e.g.\ \citet{tinney98a}, \citet{kirkpatrick99,kirkpatrick00}, 
\citet{pavlenko07}, \citet{cruz09}, and \citet{zapatero14a}.
In the T dwarf regime, only Luhman\,16B is known to harbour Li in absorption
with an equivalent width of $\sim$8\AA{} \citep{faherty14a,lodieu14c}
whereas Li is undetected in Gl\,229\,B \citep{schultz98,oppenheimer98} and in 
the two component of Epsilon Indi B \citep{king10b}. Lithium remained undetected
in the optical spectrum of another 17 T dwarfs \citep{burgasser03d,kirkpatrick08,leggett12a}.

Ultracool subdwarfs (or Population II ultracool dwarfs) are defined as metal-poor dwarfs 
with spectral types later than M7\@. They appear bluer than the main-sequence
of solar-metallicity stars due to the dearth of metals in their atmospheres 
\citep{mould76c,baraffe97}. They usually exhibit halo kinematics, including high proper
motions and large heliocentric velocities \citep{gizis97a}. They are important 
tracers of the chemical enrichment history of the Galaxy because they belong 
to the first generations of stars. The current compendium of ultracool subdwarfs 
contains approximately 50 objects, in particular 11 L-type subdwarfs 
\citep{burgasser03b,burgasser04,cushing09,sivarani09,lodieu10a,kirkpatrick10,lodieu12b,kirkpatrick14}.

%
%
\begin{table*}
 \centering
 \caption[]{Properties of the two mid-L subdwarfs presented in this paper.
Listed are the names, coordinates, spectral types \citep{burgasser07d,burgasser09a}, 
Sloan $r$-band magnitude \citep{adelman_mccarthy07,adelman_mccarthy09}, proper motions, 
distances \citep{schilbach09}, final upper limits on the Li\,I 
pseudo-equivalent widths, radial velocities, and revised space motion (this paper).
}
 \begin{tabular}{@{\hspace{0mm}}c @{\hspace{1mm}}c @{\hspace{2mm}}c @{\hspace{2mm}}c @{\hspace{2mm}}c @{\hspace{2mm}}c @{\hspace{2mm}}c @{\hspace{2mm}}c @{\hspace{2mm}}c @{\hspace{2mm}}c c@{\hspace{0mm}}}
 \hline
R.A.     &    dec        & SpT    &  SDSS$r$ & $\mu_{\alpha}cos\delta$ & $\mu_{\delta}$ & dist & EW$_{\rm Li}$ & RV & (U,V,W) \cr
 \hline
hh:mm:ss.ss & $^{\circ}$:':'' &  & mag & mas/yr & mas/yr & pc  & m\AA{} & km~s$^{-1}$ & (km/s,km/s,km/s) \cr
 \hline
12:56:37.13 & $-$02:24:52.4 & sdL3.5 & 21.82 &  $-$741.1$\pm$1.4 & $-$1002.0$\pm$1.4 & 53.3$\pm$6.8 & $<$66$\pm$27 & $-$126$\pm$10 & ($-$62$\pm$6,$-$227$\pm$37,$-$224$\pm$18) \cr
16:26:20.14 & $+$39:25:19.5 & sdL4.0 & 20.65 & $-$1374.1$\pm$1.0 &  $+$238.0$\pm$0.9 & 33.5$\pm$1.3 & $\leq$90  & $-$239$\pm$12 & ($-$166$\pm$6,$-$261$\pm$9,$-$2$\pm$11) \cr
 \hline
 \label{tab_Lithium_sdL:properties}
 \end{tabular}
\end{table*}

Because M subdwarfs are old components of our Galaxy (a few Gyr) with effective 
temperatures in the 3000--4000\,K range \citep{rajpurohit14} we do not expect 
to detect lithium in absorption in their optical spectra because all the Li 
has been burnt. However, L subdwarfs 
are very low mass objects at low metallicities and are likely to have masses close 
to the minimum needed for Li\,I burning. Lithium is expected to be present in the 
formation of subdwarfs at high abundances given its origin in Big Bang nucleosynthesis
\citep{spite82,rebolo88a,bonifacio07}.
Hence, L subdwarfs might have preserved a significant amount of primordial Li\,I 
in their atmospheres if they are substellar.
To enable the detection of the Li\,I resonance doublet at 6707.8\AA{}, it is 
advisable to focus on the coolest subdwarfs known, with effective temperatures 
below $\sim$2000--2500\,K\@. However, the faint continua of 
L subdwarfs around the Li\,I feature at 6707.8\AA{} necessitates observations of 
sufficient resolution and sensitivity to firmly establish its presence or absence. 
For depletion factors in the range 2--100, theoretical models predict
equivalent widths (EW) between 0.3\AA{} and 0.8\AA{} \citep{pavlenko95}.
Prior, low-resolution optical observations did not have sufficient signal-to-noise 
to do this 
\citep[Table \ref{tab_Lithium_sdL:Li_lowres};][]{burgasser03b,cushing09,burgasser09a,kirkpatrick10}.

In this paper we present medium-resolution optical spectra of two
mid-L subdwarfs, SDSS\,J125637.13$-$022452.4 \citep[hereafter SDSS1256; sdL3.5;][]{sivarani09,burgasser09a} 
and 2MASS\,J162620.14$+$392519.5 \citep[hereafter 2MASS1626; sdL4;][]{burgasser07b}, 
obtained with the European Southern Observatory (ESO) Very Large Telescope 
(VLT) in Paranal (Chile) and the Gran Telescopio de Canarias (GTC) in the Roque 
de Los Muchachos Observatory on La Palma (Canary Islands).
We selected these two subdwarfs because they represent the brightest of their class. 
Their properties are summarised in Table \ref{tab_Lithium_sdL:properties}.
In Section \ref{Lithium_sdL:obs}, we describe the spectroscopic observations 
conducted in service mode by ESO and GTC\@.
In Sections \ref{Lithium_sdL:RV} and \ref{Lithium_sdL:Lum_Teff}, we present refined 
radial velocities yielding improved 3D space motions as well as luminosities and 
effective temperatures for both objects, respectively. 
In Section \ref{Lithium_sdL:LiI}, we discuss our analysis of the lithium
absorption in the optical spectra of both mid-L subdwarfs.
In Section \ref{Lithium_sdL:models}, we compare our upper limits on the
pseudo-equivalent widths of the Li\,I feature with theoretical predictions
to place a limit on the mass of the subdwarfs.
In Section \ref{Lithium_sdL:Li_halo_BD} we place our results in context
and discuss the opportunity offered to detect lithium in halo brown dwarfs.

%
%
\section{Spectroscopic observations}
\label{Lithium_sdL:obs}
\subsection{GTC/OSIRIS optical spectroscopy}
\label{Lithium_sdL:obs_GTC}

We obtained optical spectroscopy of 2MASS1626 with the OSIRIS
\citep[Optical System for Imaging and low-intermediate Resolution Integrated Spectroscopy;][]{cepa00}
spectrograph mounted on the 10.4-m GTC telescope in La Palma. Observations
were conducted in February and April 2011 over two semesters under 
programmes GTC64\_10B and GTC38\_11A (PI Espinoza Contreras; Table \ref{tab_Lithium_sdL:log_obs}).

OSIRIS is equipped with two 2048$\times$4096 Marconi CCD42-82 detectors
offering a field-of-view approximately 7$\times$7 arcmin with an unbinned
pixel scale of 0.125 arcsec. We employed the R2500R grating with a slit width
of one arcsec giving a resolution of $\sim$2400 under dark or grey conditions, 
variable seeing between 0.8 and 1.6 arcsec, 
and clear skies. We obtained three exposures shifted along the slit with on-source
integrations of 1402\,s on 5 February 2011, another three with the same on-source
integrations on 12 February 2011, and two additional exposures of 1636\,s on 
13 April 2011\@. 
Bias frames, dome flat fields, and Ne, Xe, and HgAr arc lamps were observed
by the observatory staff during the afternoon preceding or following the observations.

\begin{figure*}
  \centering
  \includegraphics[width=0.48\linewidth, angle=0]{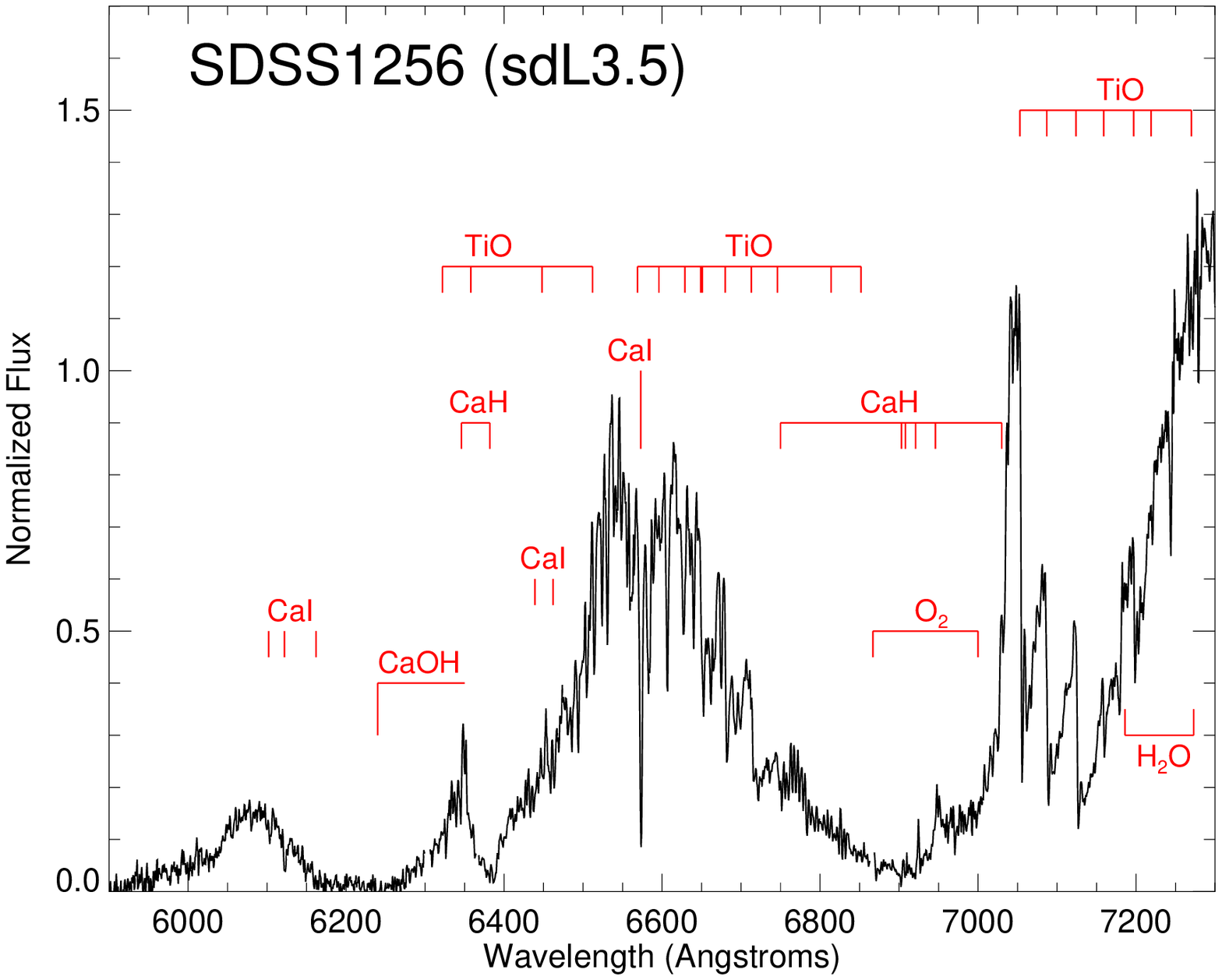}
  \includegraphics[width=0.48\linewidth, angle=0]{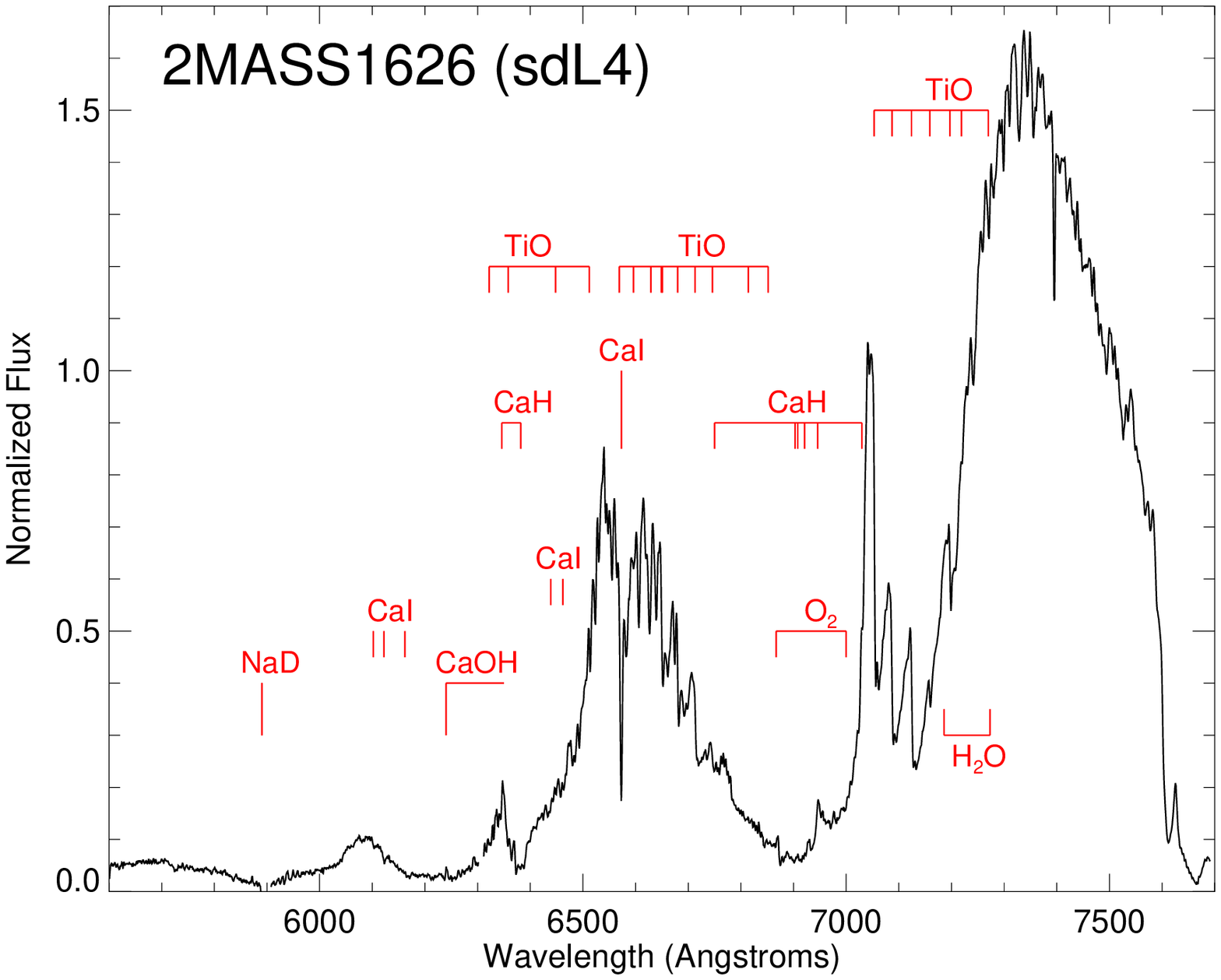}
   \caption{Full combined optical spectra of SDSS1256 (left) and 2MASS1626 (right) obtained
with VLT/FORS2 and GTC/OSIRIS, respectively. Both spectra are in the Earth frame
and have been arbitrarily normalised at 7050\AA{}. Some important bands and features 
seen in metal-poor dwarfs are marked in red.
}
   \label{fig_Lithium_sdL:full_spec_sdL}
\end{figure*}

We carried out the data reduction of the optical spectra under the IRAF environment
\citep{tody86,tody93}\footnote{IRAF is distributed by the National Optical Astronomy
Observatories, which are operated by the Association of Universities for Research in
Astronomy, Inc., under cooperative agreement with the National Science Foundation}.
We subtracted the raw median-combined bias frame from the raw spectrum and divided
by a normalised dome flat field. We extracted each spectrum individually without
combining them to avoid shifts due to the large space velocity of the target and
the position of the Earth at the time of the observations. We chose an optimal
sky background and aperture to extract the one-dimensional (1D) spectrum.
We calibrated our spectrum in wavelength with arc lamps with an accuracy better
than 0.05\AA{} rms. Finally, we calibrated each 1D spectrum in flux with the 
spectro-photometric standard stars G191-B2B \citep[DA.8;][]{vanLeeuwen07,hog00,gianninas11}
and Ross\,640 \citep[DZ5;][]{harrington80,sion09,lepine05d}. The individual
spectra were shifted by the barycentric velocities corresponding to the times
of observations, then combined with uncertainty weighting.
The final combined optical spectrum of 2MASS1626 is displayed on the right-hand
side panel in Figure \ref{fig_Lithium_sdL:full_spec_sdL}.
The signal-to-noise around the Li doublet is 90--100\@.

%
%
\begin{table}
 \centering
 \caption[]{Li\,I EWs (3$\sigma$ upper limits) from low-resolution spectra 
published in the literature, as measured here (see Section \ref{Lithium_sdL:LiI}).
References are: B03 \citep{burgasser03b}, B07 \citep{burgasser07b},
C09 \citep{cushing09}, and K10 \citep{kirkpatrick10}.
}
 \begin{tabular}{@{\hspace{0mm}}c c c c@{\hspace{0mm}}}
 \hline
Source    &  SpType &   EW(Li\,I) & Reference   \cr
 \hline
          &         &    \AA{}     &             \cr
 \hline
SDSS\,J1256$-$0122  & sdL3.5  &  $<$1.1  & B07 \\
2MASS\,J1626$+$3925 & sdL4.0  &  $<$0.4  & B07 \\
2MASS\,J0616$-$6407 & sdL5.0  &  $<$2.0  & C09 \\
2MASS\,J0532$+$8246 & sdL7.0  &  $<$0.6  & B03 \\
2MASS\,J0645$-$6646 & sdL8.0  &  $<$3.6  & K10 \\
\hline
 \label{tab_Lithium_sdL:Li_lowres}
 \end{tabular}
\end{table}
%

%
%
\begin{table}
 \centering
 \caption[]{Log of the spectroscopic observations for SDSS1256 and 2MASS1626
with VLT/FORS2 (1200R$+$93) and GTC/OSIRIS (R2500) , respectively. We give
the dates of observations and the integration times with the number of
individual exposures (in case of multiple ones) for each spectrum
obtained for each target. \newline
$^{a}$ This spectrum is not used in our analysis due to the presence of
a strong cosmic ray at the position of the Li\,I\@.
}
 \begin{tabular}{@{\hspace{0mm}}c c c c@{\hspace{0mm}}}
 \hline
Name     &  Tel.\ Inst.\  & Date     & ExpTime \cr
 \hline
         &                & dd/mm/yy & sec \cr
 \hline
SDSS1256 &  ESO VLT FORS2 & 27/04/12 & 2824 \cr
         &                & 21/06/12 & 2824 \cr
         &                & 03/04/13 & 2824$^{a}$ \cr
         &                & 11/04/13 & 2$\times$2824 \cr
         &                & 12/04/13 & 2824 \cr
         &                & 14/04/13 & 2$\times$2824 \cr
2MASS1626   &  GTC OSIRIS    & 05/02/11 & 3$\times$1682 \cr
         &                & 12/02/11 & 3$\times$1682 \cr
         &                & 13/04/11 & 2$\times$1682 \cr
 \hline
 \label{tab_Lithium_sdL:log_obs}
 \end{tabular}
\end{table}
%

%
%
\subsection{VLT/FORS2 optical spectroscopy}
\label{Lithium_sdL:obs_VLT}

We obtained intermediate-resolution (R\,$\sim$\,9000) optical spectroscopy
of SDSS1256 with the visual and near UV FOcal Reducer and low dispersion 
Spectrograph \citep[FORS2;][]{appenzeller98}
mounted on the ESO 8.2-m VLT at the Cerro Paranal observatory in Chile.
We collected eight optical spectra over two semesters under programmes
089.C-0883A and 091.C-0594A (PI Espinoza Contreras; Table \ref{tab_Lithium_sdL:log_obs})
under grey time, clear skies, and a seeing better than one arcsec.

FORS2 is a multi-mode instrument mounted on the UT1 Cassegrain focus working
at optical wavelengths. We used the standard resolution collimator providing
a pixel scale of 0.25 arcsec and a field-of-view of 6.8 arcmin by 6.8 arcmin.
We employed the grism GRIS\_1200R$+$93 associated to its order separation
filter GG435 with a slit width of one arcsec to achieve a 
spectral resolving power of $\sim$9000 around the Li\,I doublet. 
We obtained eight exposures of 2824\,s in April/June 2012 and April 2013
(Table \ref{tab_Lithium_sdL:log_obs}), yielding a total on-source integration 
of $\sim$6.3h and a signal-to-noise of 50--60 around 6700\AA{}. 
However, we do not use the spectrum taken on 03 April 2013 because the
Li\,I region is affected by a strong cosmic ray.
Bias frames, dome flat fields, and arc lamps were taken by 
the observatory staff during the afternoon.

We performed the data reduction of the FORS2 datasets in a similar way as
the OSIRIS under IRAF\@. We subtracted the raw median-combined bias frame 
from each raw spectrum, divided each of them by a normalised dome flat field, 
and extracted a 1D spectrum calibrated in wavelength to an accuracy better
than 0.05\AA{} rms. Finally, we calibrated each spectrum in flux with the
spectro-photometric standard stars LTT\,6248 \citep[A;][]{hog00,pancino12}
and LTT\,7379 \citep[G0;][]{vanAltena95,gontcharov06,vanLeeuwen07,pancino12}.
The final combined optical spectrum of SDSS1256 is shown on the left-hand side
panel in Figure \ref{fig_Lithium_sdL:full_spec_sdL}.

%
%
\begin{figure*}
  \centering
  \includegraphics[width=0.49\linewidth, angle=0]{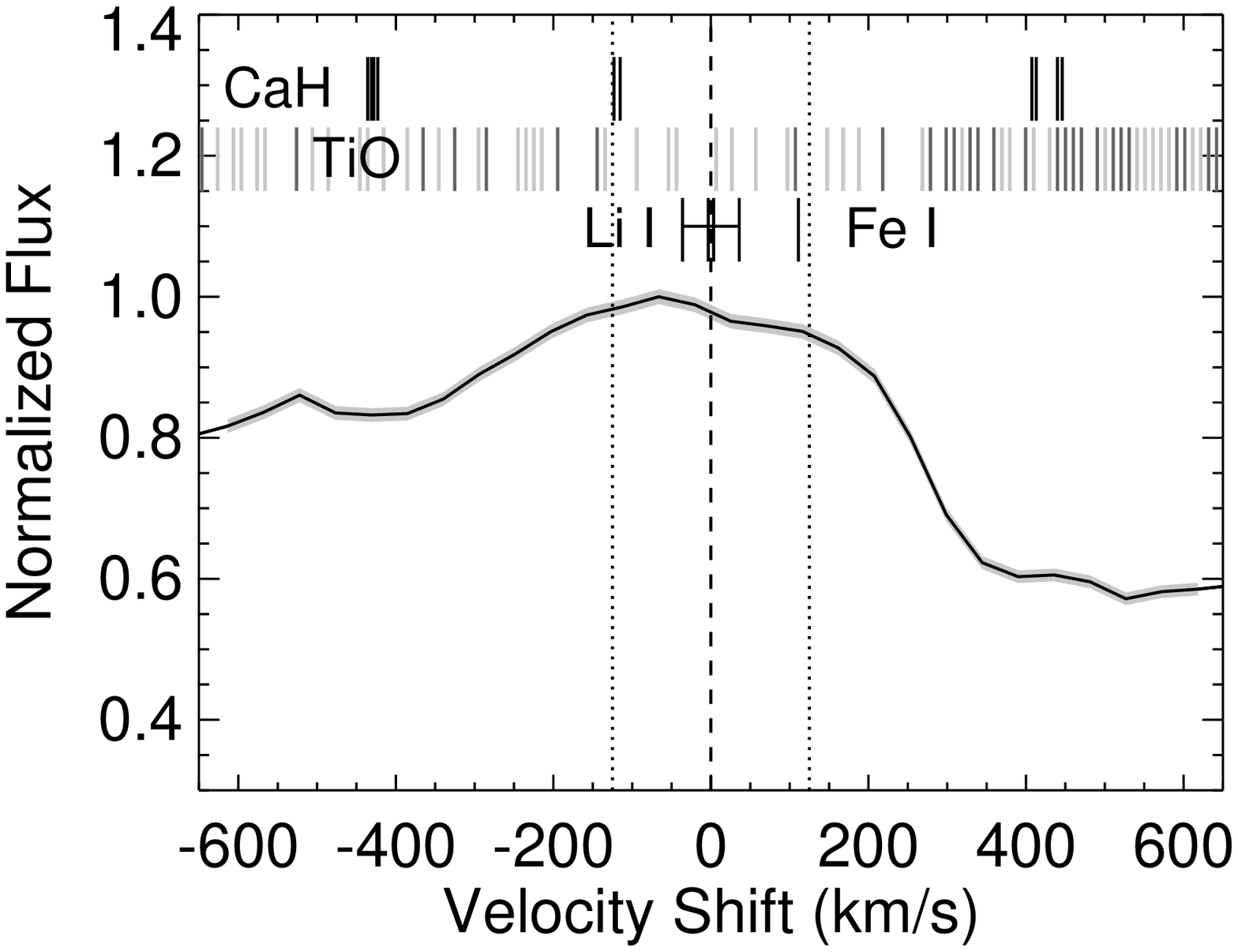}
  \includegraphics[width=0.49\linewidth, angle=0]{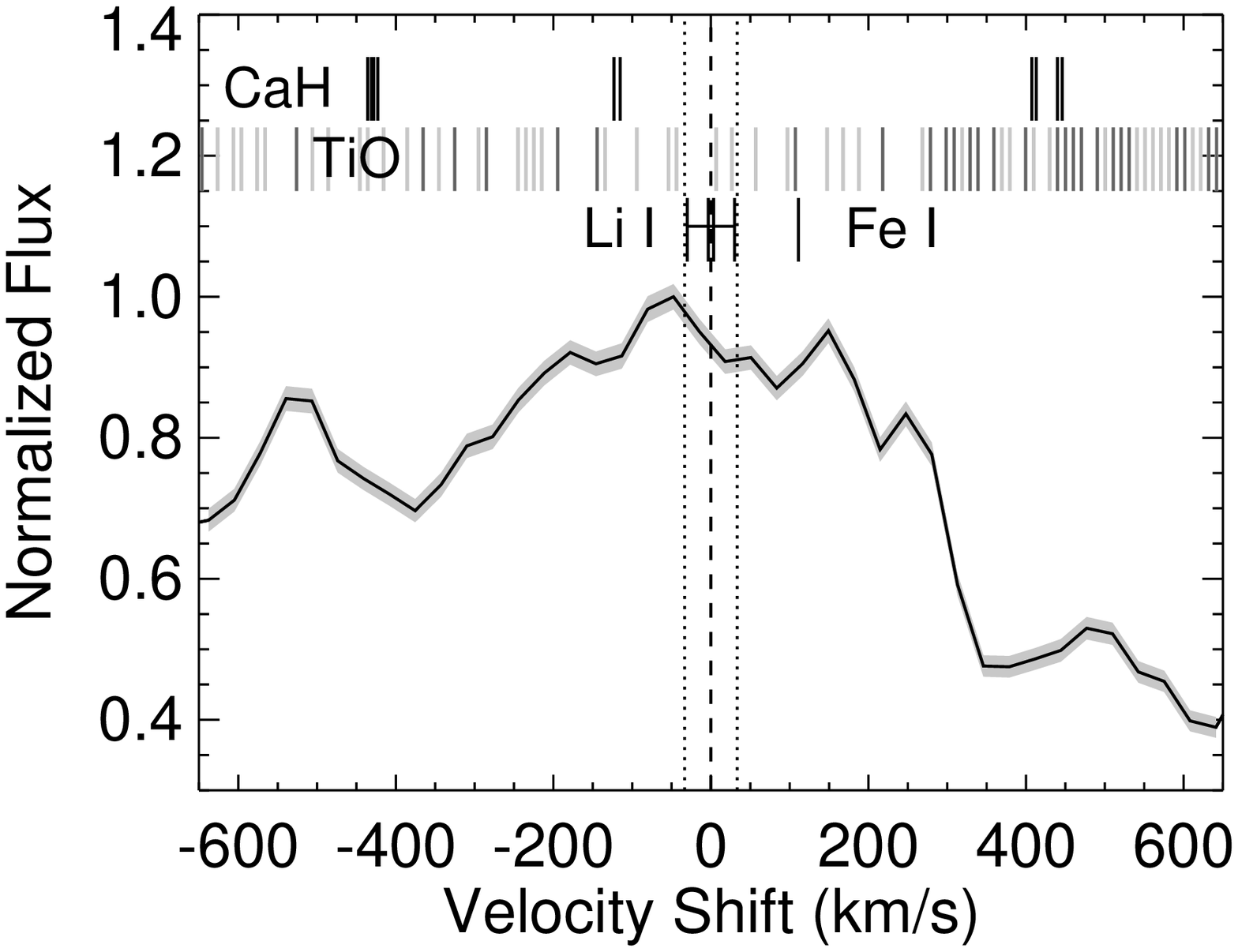}
   \caption{Zoom on the combined spectra (black line) for SDSS1256 (left) and 2MASS1626 (right)
around the Li\,I line. The grey bands indicate the formal spectral uncertainties while
the horizontal error bar indicates the 3$\sigma$ radial velocity uncertainties.
Dotted lines indicate the velocity resolution of the instruments used.
Overplotted at the top are marked the main features of CaH, Li, Fe, and TiO\@.
We note that TiO is blanketing the entire region with cross-section of various strengths 
from 5$\times$10$^{-17}$, 1$\times$10$^{-16}$, 5$\times$10$^{-16}$ cm$^{2}$/molecules
shown in three shades of grey (light to dark).
}
   \label{fig_Lithium_sdL:zoom_spec_sdL_pEWs}
\end{figure*}
%

%
%
\section{Improved radial velocities and space motions}
\label{Lithium_sdL:RV}

We took advantage of the medium resolution data for SDSS1256 and 2MASS1626
to improve their radial velocities. 
We first measured the offsets of the strong Ca{\small{I}} line at 
6572.78\AA{}\footnote{Air wavelength from the NIST database 
http://physics.nist.gov/PhysRefData/ASD/lines\_form.html},
averaging measurements from each individual spectrum.
For SDSS1256 we obtained a mean offset of 1.5$^{+0.4}_{-0.3}$\AA{}
after rejecting the value on 22 June 2012, which is discrepant by 3$\sigma$.
Assuming a spectral resolution of 0.73\AA{} equivalent to 33 km~s$^{-1}$ for
the FORS2 configuration used, we derive a mean
radial velocity of $-$152$^{+14}_{-24}$ km~s$^{-1}$, consistent with the values
of $-$98.4 km~s$^{-1}$, $-$90$\pm$40 km~s$^{-1}$, and $-$130$\pm$11 km~s$^{-1}$ 
derived by \citet{west08}, \citet{sivarani09}, and \citet{burgasser09a}, respectively. 
For 2MASS1626 we derived a mean offset of 
$-$4.60$^{+0.43}_{-0.72}$\AA{}, which translates into a radial velocity shift 
of $-$211.6$^{+19.8}_{-33.1}$ km~s$^{-1}$ assuming a spectral resolution of
1.028\AA{} for the GTC/OSIRIS configuration. This estimate agrees within
1.5$\sigma$ with the value of $-$260$\pm$35 km~s$^{-1}$ reported by \citet{burgasser04}. 

We have also measured the radial shifts by cross-correlating both spectra around the
7000--7150\AA{} TiO band to zero-velocity using SDSS templates from \citet{bochanski07a}.
We used M7, M8, and M9 templates as these provided the closest matches to the 
enhanced TiO band present in these L subdwarfs \citep[][see Section \ref{Lithium_sdL:models}]{burgasser07b}.
By correlating each of the individual spectra and averaging, we infer radial velocities 
of $-$122$\pm$9 km~s$^{-1}$ for SDSS1256 and $-$243$\pm$13 km~s$^{-1}$ for 2MASS1626,
where the uncertainties account for scatter between observations and templates. 
These velocities are in agreement with our line measurements, but with considerably
improved accuracy. We adopt as our final values the uncertainty-weighted mean of our line 
and cross-correlation measurements, which are $-$126$\pm$10 km~s$^{-1}$ for SDSS1256 
and $-$239$\pm$12 km~s$^{-1}$ for 2MASS1626\@. We also verified that these velocities aligned 
the two spectra at their rest wavelengths (Fig.\ \ref{fig_Lithium_sdL:zoom_spec_sdL_pEWs})
by checking the positions of a few lines (e.g.\ CaH at around 6698, 6705, 6717, and 6718\AA{},
Fe{\small{I}} at 6710\AA{}, and TiO at 6695, 6704.5, and 6710\AA{}) and observed structures
like to big central hump and the little bumps at $-$530 km/s and $+$450 km/s.

Moreover, we measured the pseudo-equivalent width (pEW) of the Ca{\small{I}} line at
6571.10\AA{}, inferring a mean value of 3.54$^{+0.18}_{-0.29}$\AA{} for SDSS1256 and 
3.52$^{+0.25}_{-0.40}$\AA{} for 2MASS1626, in excellent agreement, which is expected 
due to the similarity in spectral type between both targets.

We calculated UVW velocities for both sources based on our revised velocities and
reported proper motions and distances \citep{schilbach09}, assuming a local
right handed Cartesian coordinate system with U pointed radially outward from the
Galactic center, V in the direction of Galactic rotation, and W in the direction
of the Galactic North pole. We assumed a Local Standard of Rest (LSR) velocities
of (11,12.24,7.25) km~s$^{-1}$ from \citet{schoenrich10}. Our values are reported
in Table \ref{tab_Lithium_sdL:properties}. For both sources values differ from those
reported in \citet{burgasser07b,burgasser09a} primarily due to the inclusion of the 
\citet{schilbach09} astrometry. 
Both exhibit V velocities that are almost perfectly counter Galactic rotation 
at the Solar radius, 218$\pm$6 km~s$^{-1}$ \citep{bovy12a}. 

\begin{figure}
  \centering
  \includegraphics[width=\linewidth, angle=0]{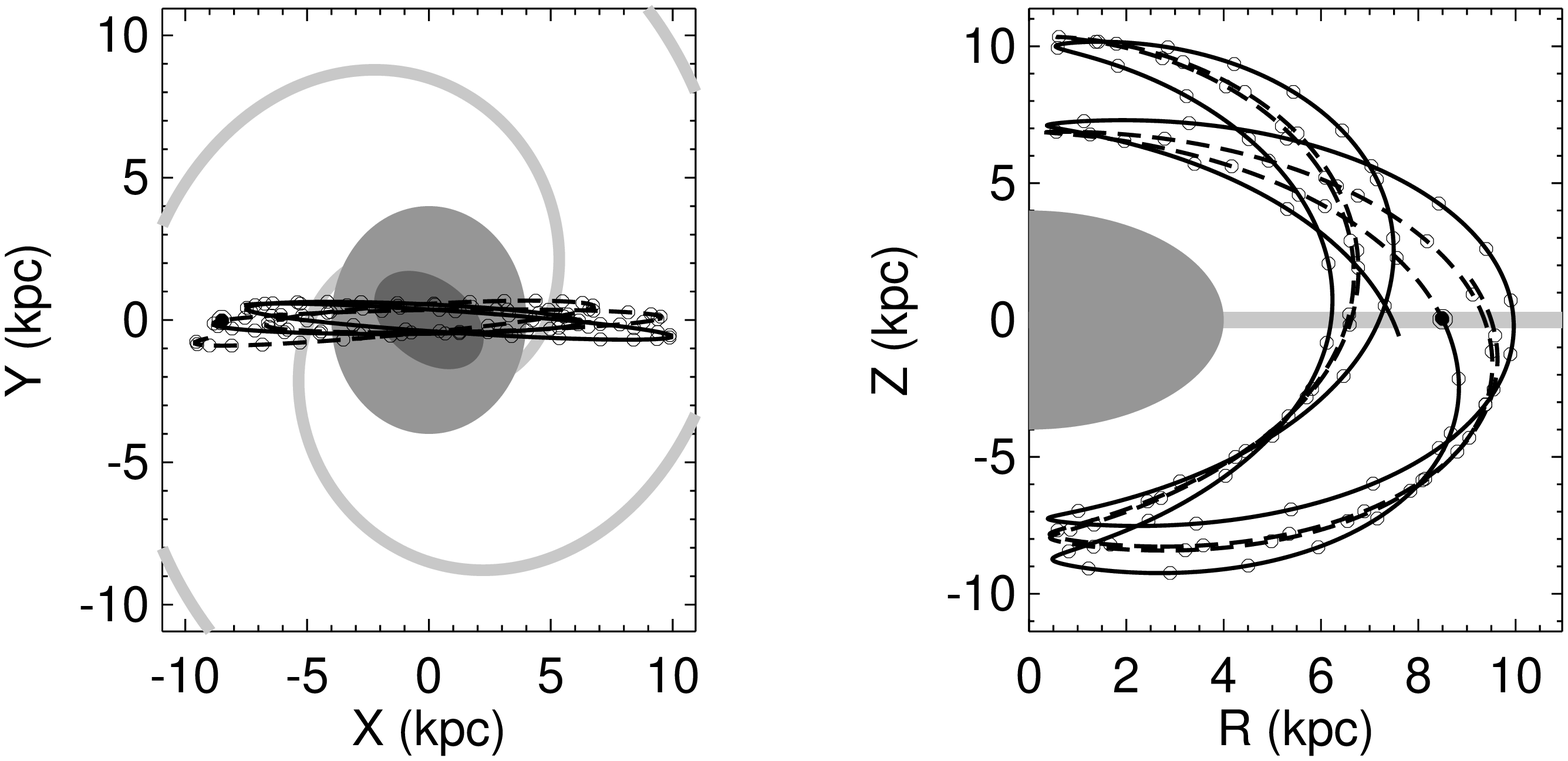}
  \includegraphics[width=\linewidth, angle=0]{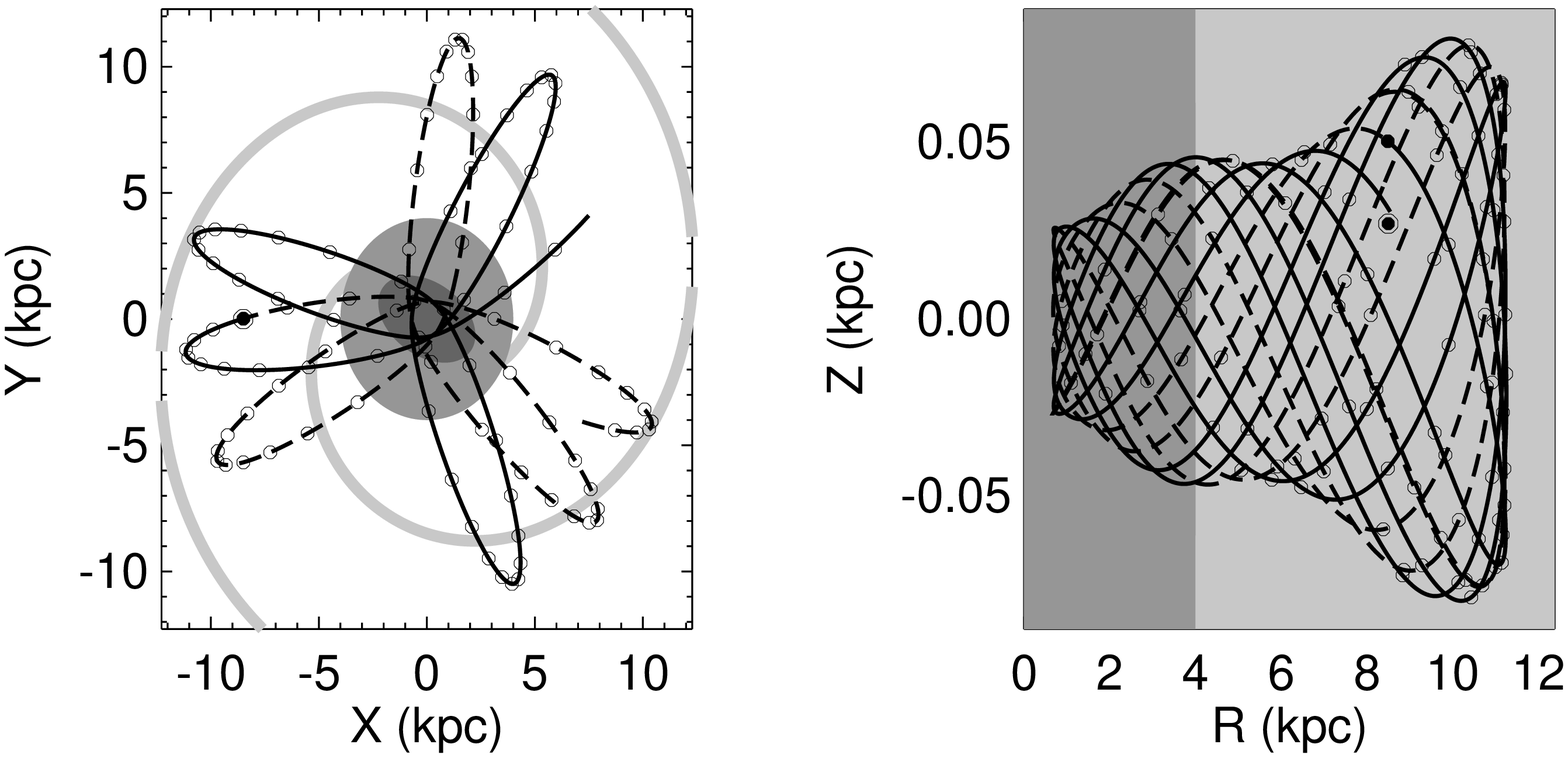}
   \caption{Galactic orbits for SDSS1256 (top) and 2MASS1626 (bottom) over
$\sim$500 Myr as viewed from above the Galactic North pole (left) and in 
cylindrical coordinates (right). future motion is indicated by solid lines, 
past motion by dashed lines, and points indicate 50 Myr steps. The Sun is
located at (X,Y)\,=\,($-$8.5,0) kpc. Representations of the Galactic bar (darkest
gray), bulge (gray), and thin disk/major spiral arms (lightest gray) are also
shown.
}
   \label{fig_Lithium_sdL:orbits}
\end{figure}

We calculated revised Galactic orbits for both sources using 
these velocities, following the methods outlined in \citet{burgasser09a}; these
are displayed in Fig.\ \ref{fig_Lithium_sdL:orbits}.
SDSS1256, whose galactocentric azimuthal velocity is nearly zero, exhibits a remarkable 
polar orbit that takes it completely around the Galactic bulge. 2MASS1626, which has a slightly
retrograde azimuthal motion and very low W velocity, remains confined to the Galactic
plane, but has an eccentric orbit that takes it within the inner kpc
of the Galaxy. Neither source goes beyond a radial
distance of 12 kpc from the Galactic center.
These orbital properties are consistent with membership to the inner halo population, 
implying a metallicity around [M/H]\,=\,$-$1.5 dex \citep{carollo07a}.

%
%
%

%
%
\section{The Li resonance doublet in mid-L subdwarfs}
\label{Lithium_sdL:LiI}

Focusing on the Li\,I region, we find that the combined spectrum of SDSS1256 shows a distinct but
weak feature near the rest wavelength of the 6707.83\AA{} Li\,I line, centered within the
radial velocity uncertainty. Using a linear fit to the adjacent peaks to estimate the
pseudo-continuum, we measure EW\,=\,66$\pm$27 m\AA{} for this feature, the uncertainty
determined from the spectral errors and radial velocity uncertainty through Monte Carlo
simulation. This 2.4$\sigma$ detection is comparable in strength to several other features
near the Li\,I line, so to assess both reliability and contamination we ompared this spectrum
to synthetic models, as discussed in Section \ref{Lithium_sdL:models}.

\begin{figure}
  \centering
  \includegraphics[width=\linewidth, angle=0]{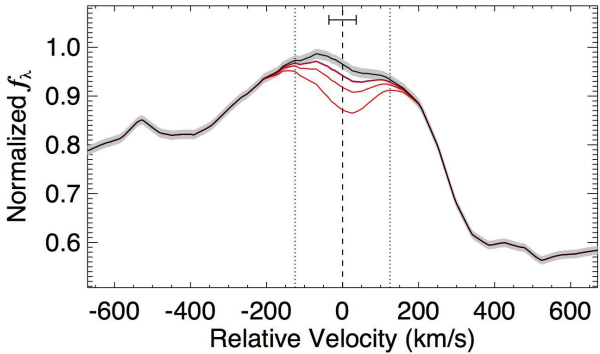}
   \caption{Zoom on the spectral region containing the Li\,I feature (black), to 
which we added Gaussian line profile (red) with EWs of 100, 200, and 400 m\AA{},
respectively.
}
   \label{fig_Lithium_sdL:model_Li_line}
\end{figure}

For 2MASS1626, no distinct features are present, largely due to the much lower resolution of the
data. To quantify our limits, we inserted a Li\,I absorption feature at progressively stronger
EWs using a Gaussian line profile modeled from the 6572.78\AA{} Ca{\small{I}} line.
Figure \ref{fig_Lithium_sdL:model_Li_line} demonstrates that statistically significant
deviations from the original spectrum occur for EW\,=\,90 m\AA{}, but a ``clear'' feature can
only be seen for an EW\,$>$\,200 m\AA{}.  We are clearly limited by systematics associated
with the low resolution of these data and the underlying structure of the pseudocontinuum.
The inferred limit is consistent with the best limits from prior low resolution observations of
L subdwarfs (Table \ref{tab_Lithium_sdL:Li_lowres}), and well above the strength of the feature
we infer for SDSS1256\@.

%
%
%

%
%
\begin{figure*}
  \centering
  \includegraphics[width=0.49\linewidth, angle=0]{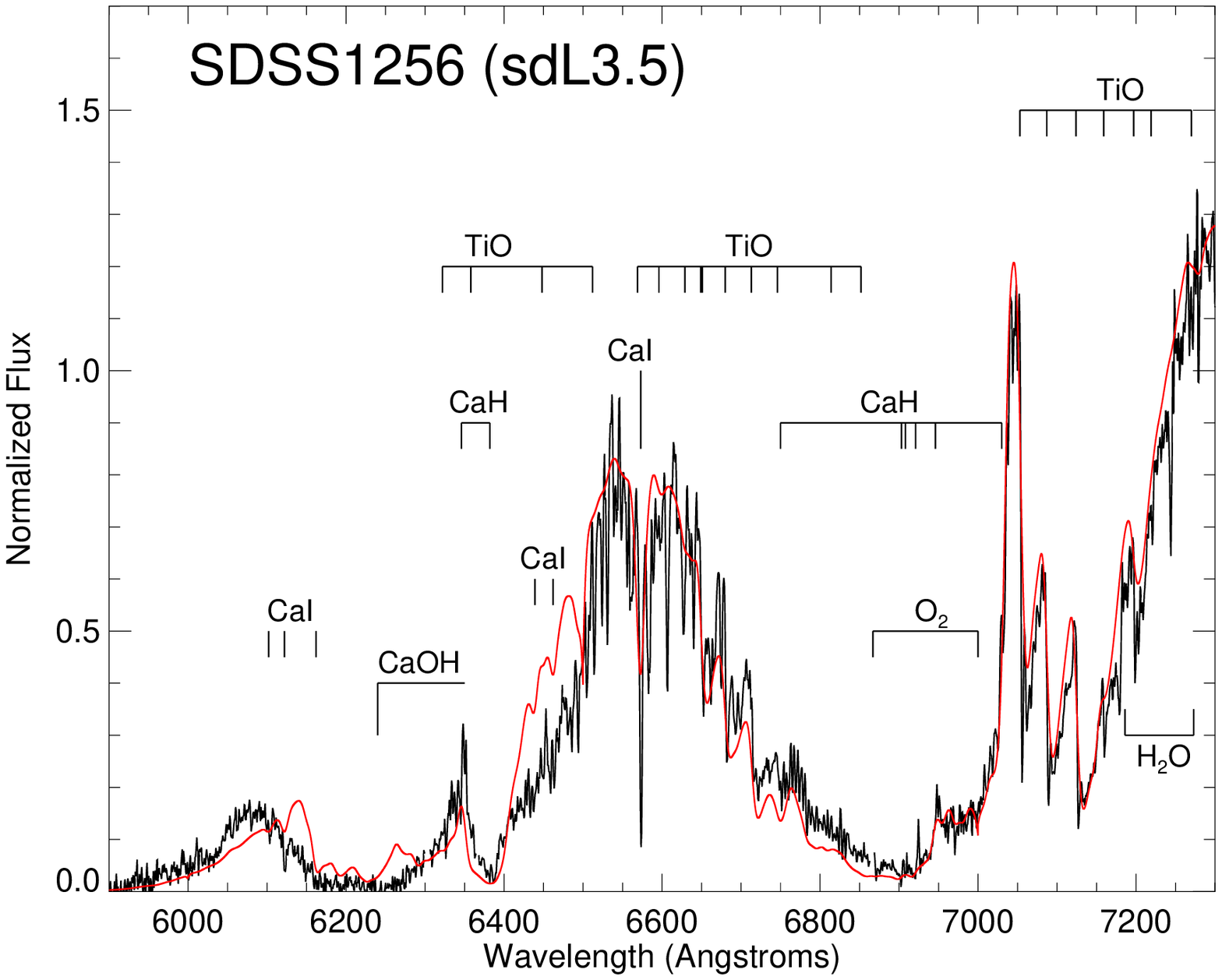}
  \includegraphics[width=0.49\linewidth, angle=0]{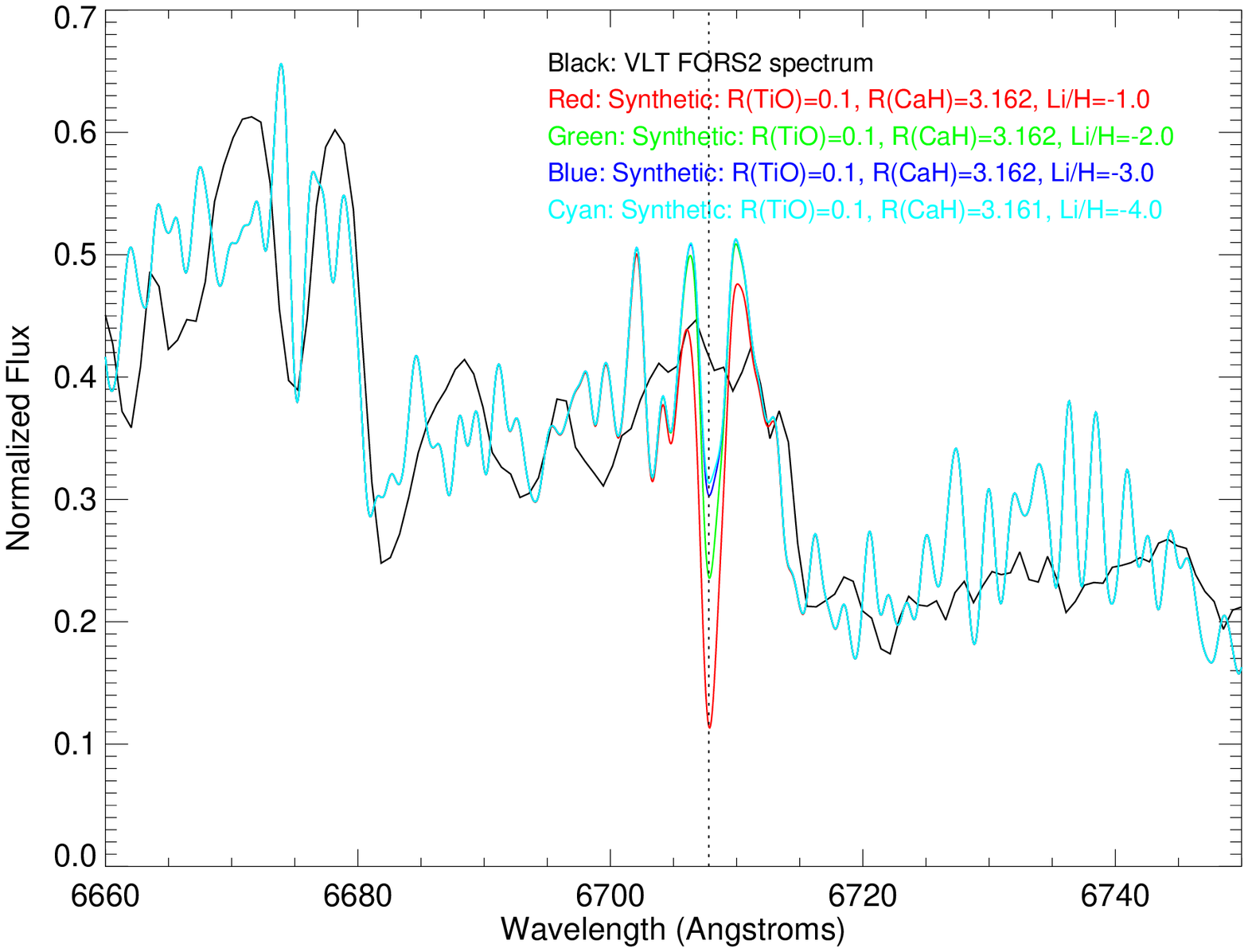}
   \caption{
{\it{Left:}} VLT/FORS2 observed spectrum of SDSS1256 (black line).
Overplotted in red is the synthetic spectrum from the NextGen atmosphere models of 
\citet{baraffe98} generated with the DUST546 code of \citet{pavlenko07}, which 
provides the best fit to the data (we show a low resolution version for plotting purposes). 
The synthetic spectra, convolved by the instrument profile, have been generated 
for a metallicity of [M/H]\,=\,$-$2.0, gravity of $\log$\,g\,=\,5.0, Li abundance 
of $\log$N(Li)\,=\,$-$1.0\@.
Both spectra are normalised to the observed spectrum around 7050\AA{}.
Major lies and features present in optical spectra of metal-poor dwarfs are
shown as well.
{\it{Right:}} Zoom on the region around the lithium doublet at 6707.8\AA{}.
The observed spectrum is shown in black while synthetic spectra with different 
Li abundances are displayed in colour.
}
   \label{fig_Lithium_sdL:spec_sdL_model}
\end{figure*}
%

%
%
\section{Comparison with theoretical models}
\label{Lithium_sdL:models}

To validate the possible presence of Li\,I in the spectrum of SDSS1256 and 
infer its atmospheric composition, we generated grids of synthetic spectra
following the procedure outlined in \citet{pavlenko07}. We started with the
NextGen model atmospheres of \citet{allard01} at solar metallicity. The 
profiles of the Na{\small{I}} and K{\small{I}} resonance doublets were
computed in the framework of a quasi-static approach described in
\citet{pavlenko07b} with an upgraded approach from \citet{burrows03}.
The line lists of VO and CaH were taken from Kurucz's
website\footnote{kurucz.harvard.edu} with more details presented in
\citet{pavlenko14}. The CrH and FeH line lists were computed by
\citet{burrows02b} and \citet{dulick03}, respectively. We upgraded the TiO
line lists of \citet{plez98} with the new version available on his
website\footnote{http://www.pages-perso-bertrand-plez.univ-montp2.fr/}.
The spectroscopic data for atomic absorption come from the
Vienna Atomic Line Database
\citep[VALD;][]{kupka99}\footnote{http://vald.astro.univie.ac.at/~vald/php/vald.php}.
More details on the technique and procedure are presented in
\citet{pavlenko06} in the case of the M6 dwarf GJ\,406\@.
Following \citet{pavlenko98a}, we computed partial pressures of some molecules
and atoms which exceeded the pressures of the gas-dust phase transition and
decreased their gas phase abundances to the corresponding equilibrium values.

\begin{figure}
  \centering
  \includegraphics[width=\linewidth, angle=0]{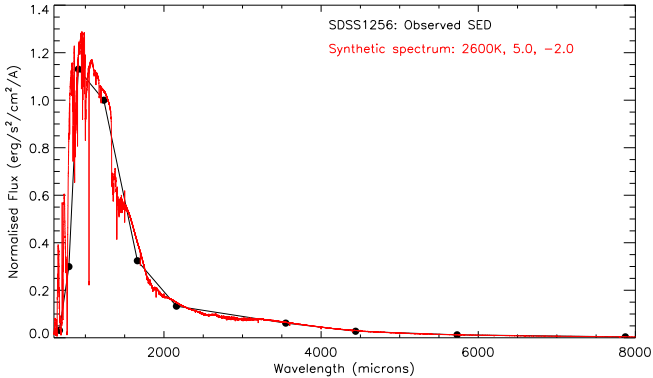}
   \caption{
Full spectral energy distribution of SDSS1256 from the optical all the way
to the mid-infrared. Overplotted in red is the synthetic spectra from the atmosphere
models of \citet{allard01} generated with the DUST546 code of \citet{pavlenko07},
which produces a satisfactory fit.
}
   \label{fig_Lithium_sdL:spec_sdL_model_full_SED}
\end{figure}

The best fit to the observed low-resolution spectrum of SDSS1256 
\citep{burgasser09a} was obtained for an effective temperature of 2600 K and a
$\log$(g) surface gravity of 5.0 (cgs). To get a better fit we reduced the 
absorption of CrH theoretical bands by factor two. This might be caused by a 
deficit of Cr atoms in the atmosphere or some depletion of Cr into dust particles. 
Then, we increased the absorption of CaH bands by factor two, likely related 
to the well known enhancement of calcium and other alpha-elements in the 
atmospheres of metal-poor dwarfs \citep{magain87a}. Moreover, we added some phenomenological 
dust opacity in the modelling of theoretical spectra, following the scheme
of \citet{pavlenko07}. The wavelength-independent dust opacity of total 
an optical depth of $\tau$\,=\,0.3 is located in a cloud-like layer at 
$\tau_{/rm ross}$\,=\,0.001, i.e above the photosphere of the dwarf. 
These phenomenologically implemented parameters provide a much improved fit to 
the spectrum of SDSS1256, reproducing the strong CaH and TiO absorption bands 
between 6300 and 7200\AA{}, as well as individual metal lines in this region
(right-hand side panel in Fig.\ \ref{fig_Lithium_sdL:spec_sdL_model}). 
The most notable discrepancy between the synthetic spectra and the observed spectrum
below 6300\AA{} is likely caused by incompleteness of the CaH linelist at these wavelengths.
The full spectral energy distribution of the SDSS1256 from the optical up to the
mid-infrared is also well-reproduced by our synthetic spectra 
(Fig.\ \ref{fig_Lithium_sdL:spec_sdL_model_full_SED}).
These adjusting parameters should be considered in the framework of more 
accurate investigation of abundances and chemical equilibrium in the atmospheres 
of metal-deficient L dwarfs because our parameters are actually not self-consistent 
with the temperature/pressure profile that generated the synthetic spectrum.

To assess the reality of the Li\,I absorption feature in the spectrum 
of SDSS1256, we examined the region around the Li\,I absorption doublet 
by fixing the adjusting factor for CaH to the aforementioned values and setting
[Li/H]\,=\,$-$1, $-$2, $-$3, and $-$4 dex (corresponding to abundances of Li
of 10, 100, 1000, and 10000 less than solar, respectively).
We note that the Li/H\,=\,$-$3.0 and $-$4.0 models are nearly identical 
because Li opacity has fallen below some other opacity source (mainly CaH).
Our observed absorption is weaker than the two most depleted abundances,
suggesting that our tentative detection is unlikely associated with
Li and instead comes from a band most likely created by CaH or TiO\@. We can
set an upper limit on the Li abundance in SDSS1256 to [Li/H]\,=\,$-$3.0\@.
We conclude that low Li abundances in subdwarfs will remain a challenge to
detect (see also Section \ref{Lithium_sdL:Li_halo_BD_accretion}).

%
%
%

%
%
\section{Luminosities and effective temperatures}
\label{Lithium_sdL:Lum_Teff}

We determined the luminosities and effective temperatures of both subdwarfs following the 
procedure described in \citet{burgasser08a}. Previously published low-resolution optical and 
near-infrared spectral data \citep{burgasser04,burgasser07b,burgasser09a} spanning 
0.6--2.5 microns were combined and scaled to 2MASS and SDSS photometry. These were joined 
to segments of a T$_{eff}$\,=\,2000\,K, $\log$\,g\,=\,5.0, solar-metallicity BT-Settl08 model 
\citep{allard12} scaled to match SDSS and WISE photometry and Spitzer/IRAC measurements for 
2MASS1626 \footnote{Measured values are [3.6]\,=\,13.27$\pm$0.04 mag, [4.5]\,=\,13.19$\pm$0.04 mag, 
[5.8]\,=\,13.03$\pm$0.06 mag, and [8.0]\,=\,12.95$\pm$0.04 mag (DDT-225, PI Burgasser).}. 
The combined spectrum was then scaled to absolute fluxes using the parallax measurements 
of \citet{schilbach09}, and integrated over 0.3--30 microns (accounting for 99.96\% of 
the total spectral model flux) to calculate bolometric fluxes and luminosities. We obtained 
radius estimates from the evolutionary models of \citet{burrows01}, \citet{baraffe03}, and
\citet{saumon08}, assuming ages of 5--10 Gyr and the measured luminosities. Values are listed 
in Table \ref{tab_Lithium_sdL:Lum_Teff_R}, with uncertainties that include spectral, photometric 
and astrometric errors, and scatter in model radii. We also experimented with different 
spectral model sets, model T$_{eff}$s spanning 1700--2400\,K, and subsolar metallicity models 
([M/H]\,=\,$-$1, $-$2); however, because the spectral model segments encompassed a small portion 
of the spectrum these variations were much smaller than the observational uncertainties. 
Indeed, varying the model parameters (i.e.\ the temperature by $\pm$500\,K and the 
gravity by $\pm$0.5 dex) did not change the resulting effective temperature by more than 15\,K\@.

Our analysis reveals that SDSS1256 is 0.2 dex less luminous and nearly 200\,K cooler than 2MASS1626, 
despite being classified 0.5 subtype earlier. This is confirmed by its fainter absolute infrared 
magnitudes ($M_{W1}$\,=\,11.49$\pm$0.30 mag vs $M_{W2}$\,=\,10.46$\pm$0.09 mag). The seeming reversal 
in temperature trend with spectral type may reflect differing metallicities between these objects
or possibly unresolved multiplicity \citep{goldman08}

%
%
\begin{table}
 \centering
 \caption[]{Bolometric luminosities ($\log_{10}(L_{bol}/L_{\odot})$), 
effective temperatures (T$_{\rm eff}$), and radii (R) for SDSS1256
and 2MASS1626\@.
}
 \begin{tabular}{@{\hspace{0mm}}c @{\hspace{2mm}}c @{\hspace{2mm}}c @{\hspace{2mm}}c@{\hspace{0mm}}}
 \hline
Name              &  $\log_{10}(L_{bol}/L_{\odot})$  & T$_{\rm eff}$ & R \cr
 \hline
                  &     dex           & K & R$_{\odot}$ \cr
 \hline
SDSS1256 (sdL3.5) &  $-$3.93$\pm$0.08 & 2000$\pm$100  & 0.090$\pm$0.003 \cr
2MASS1626   (sdL4.0) &  $-$3.73$\pm$0.02 & 2180$\pm$30  & 0.095$\pm$0.002 \cr
 \hline
 \label{tab_Lithium_sdL:Lum_Teff_R}
 \end{tabular}
\end{table}
%

%
%
\section{On the presence of lithium in halo brown dwarfs}
\label{Lithium_sdL:Li_halo_BD}
\subsection{Weak Li I in L subdwarfs from accretion}
\label{Lithium_sdL:Li_halo_BD_accretion}

Given the absence of Li\,I absorption in excess of the overlapping molecular absorption feature, 
we infer that both L subdwarfs have depleted their lithium through core fusion and have masses above 
0.065 M$_{\odot}$. Indeed, evolutionary models of \citet{baraffe97} show that 
metal-poor dwarfs with effective temperatures of 1900--2400\,K have masses
around 0.08 M$_{\odot}$ (Table \ref{tab_Lithium_sdL:Lum_Teff_R}) and have already 
depleted their lithium and lie on the main-sequence; the stellar/substellar
boundary being around 0.06 M$_{\odot}$ (with a spread of about 10\%) at 
metallicities between $-$2.0 and $-$1.0 dex.

However, it would be conceivable that a low abundance of Li could be maintained in the photosphere 
from either the intersellar medium (ISM) or circumstellar material. We can estimate the observable 
abundance in either case assuming a steady state photospheric composition, with accretion providing 
a Li source and convection and core fusion providing a Li sink:
\begin{equation}
\frac{d(f_{Li,ph}M_{ph})}{dt} \approx M_{ph}\frac{df_{Li,ph}}{dt} \approx f_{Li,acc}\frac{dM_{acc}}{dt} - f_{Li,ph}\frac{M_{ph}}{t_{dep}} = 0
\end{equation}
Here, $f_{Li,ph}$ is the steady state mass fraction of Li in the photosphere; $f_{Li,acc}$ the mass 
fraction of Li in the accreted material; $M_{ph}$ is the constant mass of the photosphere, estimated 
as 10$^{-10}M_{tot}$ $\approx$ 10$^{-11}$M$_{\odot}$ \citep{burrows93}; ${dM_{acc}}/{dt}$ is the 
instantaneous mass accretion rate in M$_{\odot}$/yr; and $t_{dep}$ is the depletion time scale for Li 
by downward convection and core burning, assumed here to be 10 Myr. These assumptions yield
\begin{equation}
\frac{f_{Li,ph}}{f_{Li,acc}} \approx \frac{N_{Li,ph}}{N_{Li,acc}} \approx \frac{t_{dep}}{M_{ph}}\frac{dM_{acc}}{dt} \approx 10^{19}\frac{dM_{acc}}{dt}  yr/M_{\odot}
\end{equation}
The ISM accretion rate can be estimated from simple ballistic accretion;\footnote{While ISM accretion 
is commonly modeled as Bondi-Hoyle accretion \citep[e.g.][]{yoshii81,frebel09}, the dynamics of a 
low-mass halo subdwarf moving through the ISM at high speed does not satisfy the conditions for this 
kind of accretion, as the dynamic scale, $GM/v^2$ $\approx$ 10$^{10}$~cm, is much smaller than the 
ISM mean free path, $(\sigma_{ISM}n_{ISM})^{-1}$ $\approx$ 3$\times$10$^{15}$~cm, assuming 
$\sigma_{ISM}$ = $\pi\times10^{-16}$~cm$^2$ and $n_{ISM}$ = 1~cm$^{-3}$ \citep{bondi44}. 
Indeed, this condition is rarely satisfied for low mass stars.} i.e., a projectile of cross section 
${\pi}R^2$ plunging through the ISM at a relative speed $v$ $\approx$ 300 km/s\footnote{The ISM is 
at rest relative to the local standard of rest, probably accurate to within +/-30 km/s
\citep[e.g.][]{frisch06}, so the relative speed is just the absolute value of (U, V, W) = 325 km/s
which we rounded down}.
In this case,
\begin{equation}
\frac{dM_{ISM}}{dt} \approx n_Hm_H\pi{R}^2v \approx N_Hm_H\pi{R}^2/t_{pass} \approx 10^{-24}~{\rm M_{\odot}/yr}
\end{equation}
where the first expression is the instantaneous accretion rate and the second is averaged over the disk 
passage, assuming $n_H$ = 1~cm$^{-3}$ and $N_H$ = 10$^{21}$~cm$^{-2}$ \citep{kalberla09}, 
$R$ $\approx$ $R_{Jup}$ = 7$\times$10$^9$~cm, and a disk passage time of 1~Myr ($W$ = 200~pc/Myr). 
For this rate:
\begin{equation}
\frac{N_{Li,ph}}{N_{Li,ISM}} \approx 10^{-5}
\end{equation}
For an Li ISM abundance consistent with primordial Big Bang nucleosysthesis
\citep[$\log{N(Li)}$\,=\,$+$2.6 or \lbrack{}Li/H\rbrack{}\,=\,1.5;][]{spergel07,lodders03} or
hot halo stars \citep[$\log{N(Li)}$\,=\,$+$2.0 or \lbrack{}Li/H\rbrack{}\,=\,0.9;][]{spite82,fields11}
this fraction corresponds to photometric abundance of [Li/H]\,=\,$-$4 to $-$3,
at or below our inferred confusion limit with CaH. Hence, it is unlikely that Li
accreted from the ISM will be in sufficient abundance to be deteced in Li-burning
L subdwarfs.

In the case of circumstellar accretion, Solar meteoritic material has a Li abundance 20 times higher 
than hot halo stars \citep[$\log{N(Li)_{ISM}}$\,=\,$+$2.2;][]{lodders03}, so the corresponding mass 
accretion rate can be 20 times lower to achieve the same basal abundance. For a detection threshold
of [Li/H]\,=\,$-$3, we can rule out circumstellar accretion onto SDSS1256 at a rate higher than
${dM_{csm}}/{dt}$ $\approx$ 6$\times$10$^{-25}$~M$_{\odot}$/yr $\approx$ 10$^{9}$~g/yr, roughly equivalent 
to the meteoric accretion rate of Earth. While not enough is known about the formation of solid bodies 
around these low mass, metal-poor stars to make a quantitative assessment, this limit suggests
very little circumstellar material remains around these sources. Note that
accretion of a larger planet could temporarily boost the lithium abundance, a mechanism that has been 
proposed for Li-rich red giants \citep{alexander67,israelian01,carlberg10}; however, such an event must 
have happened quite recently ($<$10 Myr) for an excess to be observed before the accreted Li is convected 
to the core and destroyed.

\subsection{Searches for Lithium in Cooler Subdwarfs}
\label{Lithium_sdL:Li_halo_BD_test}

Given the lack of Li in mid-type L subdwarfs,
can we confirm the substellar nature of halo brown dwarfs with the lithium test
\citep{magazzu91,rebolo92,magazzu92,basri96} in the near future?
For solar-metallicity stellar and substellar chemistry \citep[e.g.][]{burrows99,lodders06},
Li is predicted to be in atomic form above $\sim$1500--1700 K at 1 bar pressure but is 
converted into molecules at lower temperatures. Hence, we would expect chemical
depletion of Li for ultracool dwarfs below 1500\,K, which seems supported by observations
of the coolest L and T dwarfs \citep{kirkpatrick08,king10b,faherty14a,lodieu14c}.

Theoretical models locate the hydrogen-burning limit at higher masses 
($\sim$0.09 M$_{\odot}$ for $Z$\,=\,0 vs 0.075 M$_{\odot}$ for solar metallicity) and 
warmer temperatures for sub-solar metallicities \citep{burrows97,chabrier97}.
From our results for the two mid-L subdwarfs presented in this paper, we can
conclude that the Li burning limit lies below 2000\,K, equivalent to masses
below $\sim$0.08 M$_{\odot}$. The low-metallicity models of \citet{baraffe97} predict
a Li boundary minimum masses around 0.06 M$_{\odot}$ and temperatures below 1700--1600\,K
for subsolar metallicities and ages older than 1 Gyr, in the range for coolest 
L subdwarfs \citep[e.g.\ 2MASS J05325346$+$8246465;][]{burgasser08a}.
However, being substellar, the atmospheres of these objects become exceedingly
cold at the older ages expected for halo dwarfs. An object with 
[M/H]\,=\,$-$1.0 ($-$2.0) at the Li boundary mass has temperatures of 
$<$1000\,K (1050\,K) for ages larger than 5 Gyr; i.e.\ it is a T subdwarf.
If Li chemistry follows solar metallicity trends, the atomic form will likely be 
depleted in such objects. On the other hand, it is possible that the Li
condensation temperature is metallicity-dependent, just as mineral/metal condensate
formation is inferred to be from observables \citep[e.g.][]{burgasser07b} and
models \citep[e.g.][]{witte09}. The search for Li absorption in even cooler 
subdwarfs is thus critical for constraining metallicity dependencies on both
the Li-burning and hydrogen-burning mass limits, and atmospheric chemistry.

%
%
\section{Conclusions and outlook}
\label{Lithium_sdL:conclusions}

We presented optical spectra of two mid-L subdwarfs and set upper limits on the
pseudo-equivalent widths of the Li\,I doublet at 6707.8\AA{}.
We can summarise our results as follows:
\begin{itemize}
\item we report the detection of a feature in the spectrum of SDSS1256 aligned
with Li\,I at 6707.8\AA{} , but attributable to CaH\@. This sets a confusion limit 
for detectable Li abundance of [Li/H]\,=\,$-$3
\item we set an upper limit of 90 m\AA{} on the EW of the Li doublet for 
2MASS1626, placing both objects in the Li-burning regime
\item we revise previous radial velocity measurements for both objects
and derive new space motions and Galactic orbits, the latter confirming both
sources as members of the inner halo population
\item we find that ISM accretion of Li-rich material would provide a basal
abundance below our confusion limits, and constrain circumstellar accretion
to be less than 10$^{9}$ g/yr on SDSS1256
\item we reproduced the observed spectrum of SDSS1256 based on NextGen
atmosphere models with reduced metallicity and factors adjusting the amount
of TiO and CaH molecules
\end{itemize}

The next step is to attempt a 
detection in cooler L and T subdwarfs to reveal their true nature and place empirical
constraint(s) on low-metallicity evolutionary models. Further photometric searches to 
uncover T subdwarfs will likely be needed in order to investigate the presence (or absence) 
of Li in metal-poor brown dwarfs. Moreover,
additional work on the condensation of major chemistry features as a function of
temperatures and pressures in metal-poor atmospheres is needed, similar to the work of
\citet{lodders99} for solar-type brown dwarfs.

%
%
\begin{acknowledgements}
NL was funded by the Ram\'on y Cajal fellowship number 08-303-01-02\@.
This research has been supported by the Spanish Ministry of Economics and 
Competitiveness under the project AYA2010-19136\@. NL thanks Marcela Espinoza 
Contreras for her contribution during the submission of the proposals. 
The authors thank Isabelle Baraffe for her insight in the lithium boundary
mass limits. \\

AJB was a visiting professor at the Instituto de Astrof\'isica de Canarias
between September and December 2014 funded by the Tri-continental Talent programme
(CEI Canarias: Campus Atl\'antico Tricontinental). \\

This work is based on observations collected with FORS2 on the VLT at the 
European Southern Observatory, Chile, under programmes 089.C-0883 and 091.C-0594A\@.
This work is also based on observations made with the Gran Telescopio Canarias (GTC), 
installed in the Spanish Observatorio del Roque de los Muchachos of the Instituto 
de Astrof\'isica de Canarias, in the island of La Palma (programmes GTC64\_10B and GTC38\_11A).
This research has made use of the Simbad and Vizier databases, operated
at the Centre de Donn\'ees Astronomiques de Strasbourg (CDS), and
of NASA's Astrophysics Data System Bibliographic Services (ADS).

\end{acknowledgements}
%

%
%
\bibliographystyle{aa}
\bibliography{../../AA/mnemonic,../../AA/biblio_old}
%

\end{document}